\documentclass[aps,pra,twocolumn,groupedaddress,showpacs,floatfix]{revtex4-1}%
\usepackage{graphicx,amsmath,amssymb}
\usepackage{amsmath}
\usepackage{amsfonts}
\usepackage{amssymb}
\usepackage{graphicx}%
\providecommand{\U}[1]{\protect\rule{.1in}{.1in}}
\begin{document}
\title{Quantum energy teleportation in a quantum Hall system}
\author{Go Yusa}
\email{yusa@m.tohoku.ac.jp}
\affiliation{Department of Physics, Tohoku University, Sendai 980-8578, Japan}
\author{Wataru Izumida}
\affiliation{Department of Physics, Tohoku University, Sendai 980-8578, Japan}
\author{Masahiro Hotta}
\affiliation{Department of Physics, Tohoku University, Sendai 980-8578, Japan}
\date{\today
}

\begin{abstract}
We propose an experimental method for a quantum protocol termed quantum energy
teleportation (QET), which allows energy transportation to a remote location
without physical carriers. Using a quantum Hall system as a realistic model,
we discuss the physical significance of QET and estimate the order of energy
gain using reasonable experimental parameters.

\end{abstract}

\maketitle




\section{\label{sec1}Introduction}

The phenomenon of quantum teleportation (QT) has been experimentally
demonstrated in quantum optics \cite{QT,exp-QT}. But as is well known, this
protocol can teleport only \textit{information} (i.e., quantum mechanical
information or quantum states) and not \textit{physical objects}. Thus, this
protocol cannot teleport energy because that requires a physical entity to act
as an energy carrier. For example, electricity is transported over power
transmission lines by electromagnetic waves that act as the carrier. Recently,
however, one of the authors proposed a quantum protocol termed quantum
\textit{energy} teleportation (QET) that avoids the problem by using classical
information instead of energy carriers \cite{hotta}. In this counterintuitive
protocol, the counterpart of the classical \textquotedblleft transmission
line\textquotedblright is a quantum mechanical many-body system in the vacuum
state (i.e., a correlated system formed by vacuum state entanglement
\cite{R}.) The key lies using this correlated system (hereinafter, the quantum
correlation channel) to exploit the zero-point energy of the vacuum state,
which stems from zero-point fluctuations (i.e., nonvanishing vacuum
fluctuations) originating from the uncertainty principle. This energy,
however, cannot be conventionally extracted \cite{casimir} as that would
require a state with lower energy than vacuum---a contradiction. In fact, no
local operation can extract energy from vacuum, but must instead inject
energy; this property is called passivity \cite{passivity}. According to QET,
however, if we limit only the \textit{local} vacuum state instead of all the
vacuum states, the passivity of the local vacuum state can be destroyed and a
part of the zero-point energy can in fact be extracted.

As schematically illustrated in Fig. 1, a QET system to transfer energy from
subsystem A to B consists of four elements: (i)\ a quantum correlation
channel, (ii) a local measurement system for subsystem A defined on the
quantum correlation channel, (iii) a classical channel for communicating the
measurement result, and (iv) a local operation system for subsystem B.
Essentially, QET can be regarded as quantum feedback protocol implemented via
local operations and classical communication (LOCC). The procedure is as
follows: First, we measure the local field fluctuations at subsystem A. The
obtained result includes information about local fluctuations at subsystem\ B
because of the vacuum state entanglement via the quantum correlation channel
\cite{R}. This is because the kinetic energy term in the field Hamiltonian
generates the entanglement and provides partial correlation between local
vacuum fluctuations. Thus, owing to passivity of the vacuum state, the
measurement causes some energy ($E_{A}$) to be injected into subsystem A.
Next, the obtained result is communicated to subsystem B via a classical
channel. Since the measurement performed at subsystem A is \textit{local},
subsystem B remains in a local vacuum state. As mentioned above, if a good
local operation is performed at subsystem B using the information gained at
subsystem A, it will be possible to extract some amount of the zero-point
energy of subsystem B, $E_{B}$. Thus, this protocol only gives
\textquotedblleft permission\textquotedblright\ to use the otherwise
unavailable energy at B. If we define "teleportation" as a process of
transferring energy to a remote location without a physical energy carrier, we
can say that energy is teleported by this protocol.

Although the validity of this protocol has been confirmed
\textit{mathematically}, its \textit{physical} significance remains
questionable: What type of physical system is necessary for implementing QET?
What is the composition of the quantum correlation channel? Can significant
amounts of energy be \textquotedblleft teleported\textquotedblright?
Unfortunately, all past proposals for experimental verification of QET cannot
teleport sufficient amounts of energy to be measured with present technology
\cite{cold ions,EM field}. Here, we discuss a more realistic possible
implementation and estimate the order of the \textquotedblleft
teleported\textquotedblright\ energy using reasonable experimental parameters.

\begin{figure}[ptb]
\includegraphics[bb=0 0 219 124, clip,width=8.6cm]{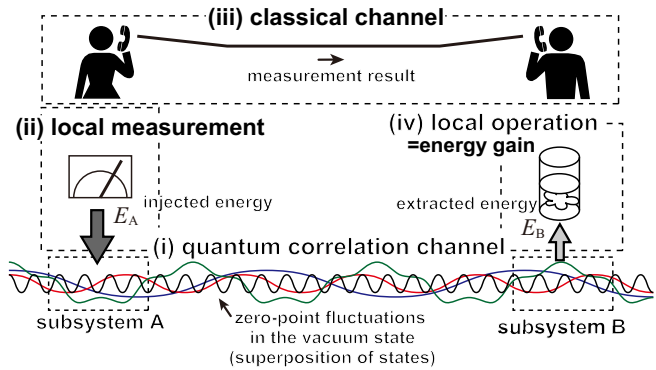} \caption{(color
online). Schematic diagram of the quantum energy teleportation (QET)
protocol.}%
\label{fig:modelQET}%
\end{figure}

\section{\label{sec2}Overview of QET protocol in the quantum Hall system}

Verification of QET in a realistic system requires the following: (i) a
dissipationless system, (ii) a quantum correlation channel with a macroscopic
correlation length, (iii) detection and operation schemes for well-defined
fluctuations in the vacuum state, and (iv) a suitable implementation of LOCC.

To this end, we consider a quantum Hall (QH) system as a potential candidate.
The QH effect is observed in two-dimensional (2D) electron systems in
semiconductors subjected to a strong perpendicular magnetic field
\cite{Yoshioka}. The QH system satisfies requirement (i) because the QH effect
does not offer any resistance. Further, in this system, quasi-one-dimensional
channels called edge channels appear at the boundary of the 2D incompressible
region of the QH system (i.e., QH bulk). Such an edge channel can behave as a
\textit{chiral} Luttinger liquid \cite{wen}, along which electric current
flows in a unidirectional manner. This attribute is indicative of the
chirality of the edge channel. Moreover, in experiments, the edge channel
shows power-law behaviors and does not have a specific decay length
\cite{chang,grayson}, preferable for fairly long-distance teleportation. Thus,
an edge channel satisfies requirement (ii). Furthermore, an edge channel can
be universally characterized by charge fluctuations described by a gapless
free boson field in the vacuum state, independently of the detailed structures
of the QH bulk state. Therefore, the target zero-point fluctuation is the
fluctuation of the charge density wave (i.e., a magnetoplasmon \cite{allen})
propagating in a unidirectional manner along an edge channel). This implies
that owing to the Coulomb interaction, a conventional capacitor can be used as
a sensitive probe and control method for detecting and manipulating zero-point
fluctuations of vacuum. Given these facts, it can be said that (iii) is
satisfied. Lastly, for a QH system, semiconductor nanotechnology can be used
to design on-chip LOCC, thus satisfying requirement (iv).

\begin{figure}[ptb]
\includegraphics[bb=0 0 218 113, clip,width=8.6cm]{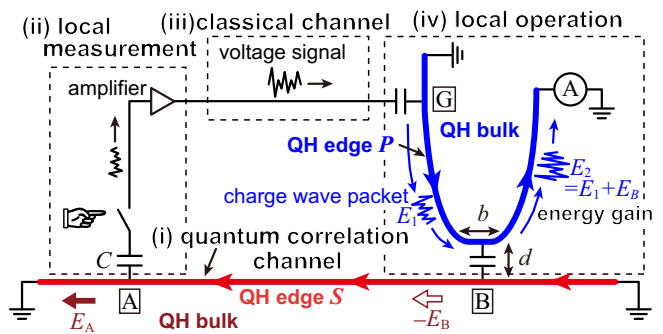} \caption{(color
online). Schematic diagram of the QH system used in this study. Edge channels
$S$ and $U$ are formed at the boundaries of QH bulk regions $S$ and $U$,
respectively. The red and blue arrows indicate the directions of propagation
of charge waves.}%
\label{fig:model}%
\end{figure}

As shown in Fig.~\ref{fig:model}, element (i), i.e., the quantum correlation
channel, is the left-going edge channel $S$. To produce the vacuum state, $S$
should be connected to an ideal electric ground, and experiments should be
performed at low temperatures---on the order of millikelvin (mK). Regions A
and B, physically corresponding to subsystems A and B, respectively, are
defined by fabricating micrometer-scale metal gate electrodes (i.e., a
microscopic capacitor) on $S$.

Element (ii), used for local measurement of the zero-point fluctuations (i.e.,
charge fluctuations), comprises a metal gate electrode fabricated on $S$ at
region A as well as an amplifier and a switch. The input resistance $R$ of the
amplifier and capacitance $C$ between $S$ and the gate electrode at region A
constitute an RC circuit. When the switch is turned on, the information on the
charge fluctuations in $S$ is imprinted on the quantum voltage fluctuations of
the electric circuit via the Coulomb interaction and then enhanced by the
amplifier. Here, the on-chip electrical circuit serves the function of the
voltmeter shown in the schematic in Fig. 1. As explained later, we can assume
that this RC circuit and amplifier can operate fast enough, and the circuit
can be considered as performing a positive operator valued measure (POVM)-type
measurement \cite{chang}. The amplified signal $\upsilon$ (i.e., measurement
result) is transferred through a classical channel (element (iii)), which
corresponds to an electric wire.

Element (iv), used for the local operation, includes another edge channel $P$
placed such that $P$ and $S$ approach each other at region B. It also consists
of a metal gate electrode fabricated on $P$ at region G and a measurement
instrument such as a picoammeter (Fig. \ref{fig:model}).

The experimental procedure is a follows: First, we cool down the entire
system, except the measurement instruments, to the lowest temperature possible
(on the order of several mK) to achieve the vacuum state. Next, we turn on the
switch only for a period of $\tau_{m}$. When a voltage signal $\upsilon$
arrives at region G, it excites a charge wave packet on $P$ via capacitive
coupling. Because of the chirality of the edge channel, the charge wave packet
travels in a unidirectional manner along $P$, carrying energy $E_{1}$ toward
region B, where the wave packet interacts with the zero-point fluctuation of
$S$\cite{chirarity}. Then, the energy carried by the wave packet changes from
$E_{1}$\ to $E_{2}$. Finally, we measure the signal with the picoammeter
connected to $P$ and thereby estimate the energy carried by the wave packet.
This is a unit cycle of a single-shot measurement. We may repeat this
single-shot measurement a sufficient number of times to generate meaningful
statistics. Finally, we can use the results to estimate the average energy
$\langle E_{2}\rangle$ carried by the wave packets. To verify that QET is
actually occurring, we must also perform a control experiment in which region
G is disconnected from the classical channel and instead connected to a signal
generator to excite wave packets \textit{independently} of $\upsilon$. If wave
packets are created by the signal generator (i.e., no information about
$\upsilon$ is communicated), they will inject energy into $S$ because of the
passivity of the vacuum state \cite{passivity}. Thus, $E_{B}=E_{2}-E_{1}$ will
be negative. However, in our system, since wave packets explicitly depend on
$\upsilon$, passivity is disturbed and $\langle E_{2}\rangle$ can take a
positive value; in other words, positive energy is extracted from the
zero-point fluctuations of $S$. Finally, if $\langle E_{2}\rangle$ is larger
in QET\ experiment than in the control, we can conclude the QET theory is
valid---energy is \textquotedblleft teleported\textquotedblright\ from A to B
without physical carriers to transport that energy. In what follows, we prove
this argument theoretically and estimate $E_{B}$ by setting the experimental
parameters $R\sim10~$k$\Omega$ \cite{impedance}; $C\sim10$~fF; and $v_{g}%
\sim10^{6}$~m/s \cite{ashoori,kamata}, were $v_{g}$ is the group velocity of a
charge density wave. Length $b$ of regions G and B and the length of region A
are approximated by a typical length scale of $l\sim10~\mu$m.

\bigskip

\section{\label{sec3}QET formulation in the QH system}

\subsection{\label{sec:level2}Formulation of chiral edge channel and local
measurement of charge fluctuations}

Here we discuss the chiral field of the edge channels. Details on the
treatment of the chiral field can be found in the literature \cite{Yoshioka}.
\ Let us start the detailed discussion with a model of the edge channel $S$.
The chiral field operator $\varrho_{S}(x)$ satisfies a commutation relation,
$\left[  \varrho_{S}(x),~\varrho_{S}(x^{\prime})\right]  =i\frac{\nu}{2\pi
}\partial_{x}\delta(x-x^{\prime})$. The energy density operator of
$\varrho_{S}(x)$ is written as%
\[
\varepsilon_{S}(x)=\frac{\pi\hbar v_{g}}{\nu_{S}}:\varrho_{S}\left(  x\right)
^{2}:,
\]
where $\nu_{S}$ is the Landau level filling factor of $S$ and $::$ denotes
normal ordering, which makes the expectation value of $\varepsilon_{S}(x)$ to
be zero for the vacuum state $|0_{S}\rangle$; $\langle0_{S}|\varepsilon
_{S}(x)|0_{S}\rangle=0$. The free Hamiltonian of $S$ is given by $H_{S}%
=\int_{-\infty}^{\infty}\varepsilon_{S}(x)dx$. The eigenvalue of the vacuum
state vanishes: $H_{S}|0_{S}\rangle=0$. If the vacuum state is not entangled,
the two-point correlation function of $\langle0_{S}|\varrho_{S}(x_{A}%
)\varrho_{S}(x_{B})|0_{S}\rangle$ with $x_{A}\neq x_{B}$ is exactly zero.
However, this entangled vacuum state provides a nontrivial correlation:
\[
\langle0_{S}|\varrho_{S}(x_{A})\varrho_{S}(x_{B})|0_{S}\rangle=-\frac{\nu_{S}%
}{4\pi^{2}\left(  x_{A}-x_{B}\right)  ^{2}}.
\]
This correlation function can be calculated by using creation and annihilation
operators of the free field. Taking region A for $x\in\left[  a_{-}%
,~a_{+}\right]  $, we adopt the RC-circuit-detector model proposed by F\`{e}ve
\textit{et al}. \cite{Feve} to measure the voltage induced by the zero-point
fluctuations of $\varrho_{S}(x)$. The charge fluctuation at A is estimated as
\begin{equation}
Q_{S}(t)=e\int_{-\infty}^{\infty}\varrho_{S}\left(  x+v_{g}t\right)
w_{A}(x)dx \label{v}%
\end{equation}
with a window function $w_{A}(x)$ that equals 1 in $x\in\left[  a_{-}%
,~a_{+}\right]  $ and decays rapidly outside A. In this model \cite{Feve}, the
voltage at the contact point between the detector and $S$ is given by
$V(t)=\frac{1}{C}\left[  Q_{S}(t)-Q(t)\right]  $, where $Q(t)$ is the charge
of the capacitor. The coupled Hamiltonian of $S$ and the RC circuit can be
directly diagonalized, enabling analytical estimation of various physical
quantities \cite{Feve}. For example, the quantum noise of the voltage $V(t)$
is described by an operator $\hat{V}$ defined by%
\begin{align}
\hat{V}  &  =-\sqrt{\frac{\hbar}{\pi RC^{2}}}\nonumber\\
&  \times\int_{0}^{\infty}d\omega\left[  \frac{\sqrt{\omega}}{\omega-\frac
{1}{iRC}}a_{in}(\omega)+\frac{\sqrt{\omega}}{\omega+\frac{1}{iRC}}a_{in}%
^{\dag}(\omega)\right]  , \label{f1}%
\end{align}
where $a_{in}(\omega)$($a_{in}(\omega)^{\dag}$) is the annihilation (creation)
operator of excitation of the charge density wave in the local-measurement RC
circuit and $\left[  a_{in}(\omega),~a_{in}(\omega^{\prime})^{\dag}\right]
=\delta\left(  \omega-\omega^{\prime}\right)  $. Prior to the measurement
(i.e., the signal input from $S$ to the detector), $V(t=-0)$ equals $\hat{V}$.
Using the fast detector condition ($RC\ll l/v_{g}$), the voltage after the
measurement is computed as
\begin{equation}
V(t=+0)=\hat{V}+R\dot{Q}_{S}(0), \label{1}%
\end{equation}
where $R\dot{Q}_{S}(0)$ denotes the voltage shift induced by the signal and
the dot in $\dot{Q}_{S}(0)$ stands for the time derivative. Using Eq.
(\ref{f1}), the amplitude $\Delta V$\ of $\hat{V}$ in the vacuum state
$|0_{RC}\rangle$ of the RC circuit can be estimated as%
\[
\Delta V=\sqrt{\langle0_{RC}|\hat{V}^{2}|0_{RC}\rangle}\sim\sqrt{\frac{\hbar
}{RC^{2}}},
\]
which is expected to be on the order of $10$~$\mu$V. From Eq.(\ref{v}), the
root-mean-square value of the voltage sift,$\sqrt{\langle0_{S}|\left(
R\dot{Q}_{S}(0)\right)  ^{2}|0_{S}\rangle}$, is estimated to be on the order
of $100~\mu$V, showing that the quantum fluctuations of the edge current are detectable.

Now, we estimate the corresponding measurement operators \cite{nc} of this
voltage measurement. Clearly, this is difficult to achieve with sufficient
accuracy with a microscopic model. However, after the amplification of the
quantum noise of the voltage $V(t)$, the signal becomes macroscopic and
classical. Thus, we may estimate the measurement operators of the macroscopic
system comprising subsystem A, the amplifier, and the electric wire by
reducing the measurement to the pointer-basis proposed by von Neumann
\cite{vN}. For this, let us begin with a gedankenexperiment in which a
high-speed voltage meter is connected to the amplifier. Thus, the position of
the meter pointer instantaneously shifts according to the signal strength.
Assume that the pointer shift is equal to Eq. (\ref{1}). In the same manner as
that used by von Neumann \cite{vN}, we can treat the macroscopic system
including this voltage meter with quantum mechanics, even though the meter is
macroscopic and classical. The readout of the meter pointer can be, therefore,
treated as a kind of quantum measurement, which can be described by
measurement operators $M_{v}$ \cite{nc} with the output value of $v$. The
shift of the meter pointer, $R\dot{Q}_{S}(0)$, in Eq. (\ref{1}) can be
reproduced by a macroscopic measurement Hamiltonian given by%

\[
H_{m}(t)=\hbar\delta(t)R\dot{Q}_{S}(0)P_{\hat{V}},
\]
where $P_{\hat{V}}$ is the conjugate momentum operator of $\hat{V}$. In fact,
the time evolution generated by this effective Hamiltonian is given by\textbf{
}%
\[
U_{m}=\operatorname*{T}\exp\left(  -\frac{i}{\hbar}\int_{-0}^{+0}%
H_{m}(t)dt\right)  =\exp\left(  -iR\dot{Q}_{S}(0)P_{\hat{V}}\right)
\]
with time-ordered exponentiation, $\operatorname*{T}\exp$, of the time
dependent Hamiltonian and reproduces Eq. (\ref{1}) as follows:
\[
U_{m}^{\dag}\hat{V}U_{m}=V(t=+0)=\hat{V}+R\dot{Q}_{S}(0).
\]
We are able to derive the measurement operators $M_{v}$ by using $U_{m}$.
Firstly, using the eigenvalue $\upsilon$ of $\hat{V}$ ($\hat{V}|\upsilon
\rangle=\upsilon|\upsilon\rangle$), we can assume the initial wavefunction of
the quantum pointer in the $\upsilon$ representation as
\[
\Psi_{i}(\upsilon)\varpropto\exp\left[  -\frac{1}{4\Delta V^{2}}\upsilon
^{2}\right]  ,
\]
whereas the wavefunction after the measurement is translated as
\[
\Psi_{f}(\upsilon)\varpropto\exp\left[  -\frac{1}{4\Delta V^{2}}\left(
\upsilon-R\dot{Q}_{S}(0)\right)  ^{2}\right]  ,
\]
using $U_{m}$. Next, after turning the measurement interaction on (i.e.,
turning the switch on), we perform a projective measurement of $\hat{V}$ to
obtain an eigenvalue $\upsilon$ of $\hat{V}$. This reduction analysis proves
the measurement operator $M_{\upsilon}$ being $\Psi_{f}(\upsilon)$:%
\[
M_{\upsilon}=\left(  \frac{1}{2\pi\Delta V^{2}}\right)  ^{1/4}\exp\left[
-\frac{1}{4\Delta V^{2}}\left(  \upsilon-R\dot{Q}_{S}(0)\right)  ^{2}\right]
.
\]
The corresponding POVM is given by $\Pi_{\upsilon}=M_{\upsilon}^{\dag
}M_{\upsilon}$ and satisfies the standard sum rule: $\int_{-\infty}^{\infty
}\Pi_{\upsilon}d\upsilon=I_{S}$, where $I_{S}$ is the identity operator of the
Hilbert space of $\varrho_{S}(x)$. The emergence probability density of the
result being $\upsilon$ is $p(\upsilon)=\langle0_{S}|\Pi_{\upsilon}%
|0_{S}\rangle$. The post-measurement state of $\varrho_{S}\left(  x\right)  $
corresponding to the result $\upsilon$ is computed as $M_{\upsilon}%
|0_{S}\rangle$ up to the normalization constant. Hence, the average state of
$\varrho_{S}\left(  x\right)  $ right after the measurement is given by
\[
\rho_{1}=\int_{-\infty}^{\infty}M_{\upsilon}|0_{S}\rangle\langle
0_{S}|M_{\upsilon}^{\dag}d\upsilon.
\]
The amount of energy injected by the measurement is calculated as%
\begin{align*}
E_{A}  &  =\int_{-\infty}^{\infty}\langle0_{S}|M_{\upsilon}^{\dag}%
H_{S}M_{\upsilon}|0_{S}\rangle d\upsilon\\
&  =\frac{\hbar v_{g}\nu_{S}}{4\pi}\left(  \frac{ev_{g}R}{2\Delta V}\right)
^{2}\int_{-\infty}^{\infty}dx\left(  \partial_{x}^{2}w_{A}(x)\right)  ^{2}.
\end{align*}
Using the experimental parameters mentioned earlier, $E_{A}$ can be estimated
to be on the order of $1~$meV for $\nu_{S}\sim3$. Since the meter we consider
is sufficiently macroscopic such that quantum effects can be neglected, the
estimation of $M_{\upsilon}$ and $E_{A}$ remains unchanged even if we directly
send the amplified classical signal to region G without the voltage meter we
assumed above.

\subsection{\label{sec:level2 2}Formulation of local operation and estimated
energy gain at B\bigskip}

Now, let us turn to the edge channel $P$ and discuss how wave packets can be
excited at G (i.e., how to send the measurement result to B). After the
measurement result $\upsilon$ is amplified and transferred to region G as a
voltage signal through the wire, the voltage signal (i.e., the electric field)
excites a charge wave packet of $\varrho_{P}(y)$. Here, $\varrho_{P}(y)$ is
the chiral field operator, the counterpart of $\varrho_{S}(x)$ in the edge
channel $S$. In other words, by performing a $\upsilon$-dependent unitary
operation $U_{\upsilon}$ on the vacuum state $|0_{P}\rangle$ of $\varrho
_{P}(y)$, a localized right-going coherent state is generated: $|\upsilon
_{P}\rangle=U_{\upsilon}|0_{P}\rangle$ in\ a region with $y\in\left[
b_{-}-L,~b_{+}-L\right]  $, where $L$ is the distance between regions G and B.
The length $b_{+}-b_{-}\,$of region B is given by $b$ $(\sim l)$. This
operation is realized by applying an electric field with a strength
proportional to the measurement $\upsilon$ on the edge channel $P$. Such a
unitary operation is experimentally feasible, since charge coherent states
have been demonstrated experimentally in semiconductor quantum dots
\cite{hayashi}. However, in order to realize QET experimentally, proper tuning
of the unitary operation $U_{\upsilon}$ is important. Here, let $F_{\upsilon
}(y,t)$ be the electric potential (i.e., classical external potential)
produced by the amplified voltage signal at region G. By using $F_{\upsilon
}(y,t)$, the interaction Hamiltonian of $U_{\upsilon}$ is given by a linear
term of $\varrho_{P}\left(  y\right)  $ as%

\begin{equation}
H_{\upsilon}=\int_{b_{-}-L}^{b_{+}-L}F_{\upsilon}(y,t)\varrho_{P}\left(
y\right)  dy, \label{int}%
\end{equation}
Taking negative values of $F_{\upsilon}(y,t)$ ensures that the sign of $E_{B}$
is positive. A standard inverting amplifier allows us to achieve this sign
reversal for $F_{\upsilon}(y,t)$ with respect to $\upsilon$. Now, we assume
the potential $F_{\upsilon}(y,t)$ is\ as follows and we, then, discuss how to
generate this potential experimentally.
\[
F_{\upsilon}(y,t)=-\frac{\pi\hbar}{\nu_{P}\Delta V}\upsilon\lambda
_{B}(y)\delta_{\tau_{m}}(t-t_{o}),
\]
where $\delta_{\tau_{m}}(t-t_{o})$ is a real localized function at $t_{o}$
with a short-time width $\tau_{m}$ satisfying $\lim_{\tau_{m}\rightarrow
0}\delta_{\tau_{m}}(t-t_{o})=\delta(t-t_{o})$. In addition, $\lambda_{B}(y)$
is a window function related to the total number of excited electrons and
quasi-holes from the vacuum state. In other words, the excited wave packet,
which extends over the region with $\left[  b_{-}-L,~b_{+}-L\right]  $,
contains the same order of $\lambda_{B}(y)$. Therefore, $\lambda_{B}(y)$ is
related to the shape of the metal gate electrode at region G. By using
$\left[  \varrho_{P}(y),~\varrho_{P}(y^{\prime})\right]  =-i\frac{\nu_{P}%
}{2\pi}\partial_{y}\delta(y-y^{\prime})$, the wave form is computed as
$\langle\upsilon_{P}|\varrho_{P}(y)|\upsilon_{P}\rangle=\frac{\upsilon
}{2\Delta V}\partial_{y}\lambda_{B}(y)$. Because the charge density$\ \langle
\upsilon_{P}|\varrho_{P}(y)|\upsilon_{P}\rangle$ can be directly measured in
experiments, $\lambda_{B}(y)$ is also measured depending on the design of the
gate electrode at G. Here, we take the amplitude of $\lambda_{B}(y)$ to be on
the order of $10$. To clarify the relation between $F_{\upsilon}(y,t)$ and the
voltage signal $\upsilon$, let us analyze the gain $\alpha$ of the amplifier.
By setting $\alpha$ as%
\[
\alpha=\frac{\pi\hbar}{\nu_{P}\Delta V\tau_{m}}\max_{y}\lambda_{B}(y),
\]
the potential $F_{\upsilon}(y,t)$ is order-estimated as
\[
O(F_{\upsilon})=\alpha O(\upsilon)=\alpha\Delta V.
\]
This suggests that the order of the potential is simply proportional to the
quantum noise $\Delta V$ multiplied by the gain. If $\tau_{m}$ is of
nanosecond order, $F_{\upsilon}(y,t)$ is on the order of $10$~$\mu$V. The
amplitude and the spatial profile of $F_{\upsilon}(y,t)$ is, thus,
experimentally tunable by the gain of the amplifier and the shape of the gate
electrode, respectively. Using the approximation $\tau_{m}\sim0$, this simple
interaction in Eq. (\ref{int}) generates a displacement operator given by
\[
U_{\upsilon}=\exp\left(  \frac{\pi i\upsilon}{\nu_{P}\Delta V}\int_{b_{-}%
-L}^{b_{+}-L}\lambda_{B}(y)\varrho_{P}\left(  y\right)  dy\right)  .
\]
The composite state of $S$ and $P$ at a time $T$, when generation of a charge
wave packet completes, is calculated as
\[
\rho_{SP}=\int_{-\infty}^{\infty}d\upsilon e^{-\frac{iT}{\hbar}H_{S}%
}M_{\upsilon}|0_{S}\rangle\langle0_{S}|M_{\upsilon}^{\dag}e^{\frac{iT}{\hbar
}H_{S}}\otimes|\upsilon_{P}\rangle\langle\upsilon_{P}|.
\]
This state is the scattering input state for the Coulomb interaction between
$S$ and $P$. Then, the charge wave packet evolves into region B by the free
Hamiltonian,
\[
H_{B}=\frac{\pi\hbar v}{\nu_{P}}\int_{-\infty}^{\infty}:\varrho_{P}\left(
y\right)  ^{2}:dy.
\]
The average value of the energy of the wave packet $E_{1}=\operatorname*{Tr}%
\left[  H_{B}\rho_{SP}\right]  $. This is calculated as
\[
E_{1}=\frac{\pi\hbar v_{g}}{\nu_{P}}\int_{-\infty}^{\infty}\left(
\partial_{y}\lambda_{B}(y)\right)  ^{2}dy\left[  \langle0_{S}|G_{S}^{2}%
|0_{S}\rangle+\frac{1}{4}\right]  ,
\]
where
\[
G_{S}=-\frac{ev_{g}R}{2\Delta V}\int_{-\infty}^{\infty}\varrho_{S}\left(
x\right)  \partial_{x}w_{A}(x)dx.
\]
Here, $E_{1}$ is estimated to be on order of $10~$meV for $\nu_{S}$ and
$\nu_{P}$ of $3$ and $6$, respectively. At region B, the two channels $S$ and
$P$ interact with each other via Coulomb interaction such that
\[
H_{int}=\frac{e^{2}}{4\pi\epsilon}\int_{b_{-}}^{b_{+}}dx\int_{b_{-}}^{b_{+}%
}dy\varrho_{S}\left(  x\right)  f(x,y)\varrho_{P}\left(  y\right)  .
\]
Here, $\epsilon$ is $10\epsilon_{0}$ for the host semiconductor (e.g., gallium
arsenide, GaAs), where $\epsilon_{0}$ is the dielectric constant of vacuum.
The function $f(x,y)$ is given by $\frac{1}{\sqrt{\left(  x-y\right)
^{2}+d^{2}}}$, and $d$ ($\sim l$) is the separation length between the two
edge channels at B. After exchanging energy with $\varrho_{S}\left(  x\right)
$, the energy carried by the wave packet becomes $E_{2}$. The energy gain,
$E_{B}=E_{2}-E_{1}$, is estimated by the lowest-order perturbation theory in
terms of $H_{int}$ as follows:%
\begin{align*}
E_{B}  &  =-i\frac{e^{2}v_{g}}{4\epsilon\nu_{S}}\int_{-\infty}^{\infty}%
dz\int_{b_{-}}^{b_{+}}dx_{B}\int_{b_{-}}^{b_{+}}dy_{B}f(x_{B},y_{B})\\
&  \times\int_{-\infty}^{\infty}dt\int_{-\infty}^{\infty}d\upsilon\langle
0_{S}|M_{\upsilon}^{\prime\dag}\varrho_{S}\left(  x_{B}+v_{g}t\right)
M_{\upsilon}^{\prime}|0_{S}\rangle\\
&  \times\langle\upsilon_{P}^{\prime}|\left[  \varrho_{B}\left(  z-v_{g}%
t_{f}\right)  ^{2},~\varrho_{B}\left(  y_{B}-v_{g}t\right)  \right]
|\upsilon_{P}^{\prime}\rangle,
\end{align*}
where $M_{\upsilon}^{\prime}=U_{S}(t_{i}-T)^{\dag}M_{\upsilon}U_{S}(t_{i}-T)$
and $|\upsilon_{P}^{\prime}\rangle=U_{B}(t_{i})^{\dag}|\upsilon_{B}\rangle$.
By substituting the commutation relation given by $\left[  \varrho_{B}\left(
z\right)  ^{2},~\varrho_{B}\left(  y_{B}\right)  \right]  =-i\frac{\nu_{S}%
}{\pi}\partial\delta(z-y_{B})\varrho_{B}\left(  z\right)  $ and performing the
$z$ integral, we obtain the following relation:
\begin{align*}
E_{B}  &  =\frac{e^{2}v_{g}}{4\pi\epsilon}\int_{b_{-}}^{b_{+}}dx_{B}%
\int_{b_{-}}^{b_{+}}dy_{B}f(x_{B},y_{B})\\
&  \times\int_{-\infty}^{\infty}dt\partial^{2}\lambda_{B}(y_{B}-v_{g}\left(
t-t_{i}\right)  )\\
&  \times\int_{-\infty}^{\infty}\upsilon\langle0_{S}|M_{\upsilon}^{\prime\dag
}\varrho_{S}\left(  x_{B}+v_{g}t\right)  M_{\upsilon}^{\prime}|0_{S}\rangle
d\upsilon.
\end{align*}
Note that the last integral is computed as

\begin{align*}
&  \int_{-\infty}^{\infty}\upsilon\langle0_{S}|M_{\upsilon}^{\prime\dag
}\varrho_{S}\left(  x_{B}+v_{g}t\right)  M_{\upsilon}^{\prime}|0_{S}\rangle
d\upsilon\\
&  =-\frac{e\upsilon R}{4\Delta V}\int_{-\infty}^{\infty}d\bar{x}_{A}\partial
w_{A}(\bar{x}_{A})\\
&  \times\Delta\left(  \bar{x}_{A}-x_{B}-v_{g}(t+T-t_{i})\right)  +c.c.,
\end{align*}
where
\[
\Delta\left(  x\right)  =\frac{\nu_{S}}{4\pi^{2}}\int_{0}^{\infty}%
dkk\exp\left(  -ikx\right)  .
\]
For the $t$ integral of $E_{B}$, let us use the Fourier transform
$\partial^{2}\lambda_{B}$ in $E_{B}$ as
\[
\partial^{2}\lambda_{B}(y)=-\frac{1}{2\pi}\int_{-\infty}^{\infty}k^{\prime
2}\tilde{\lambda}_{B}(k^{\prime})e^{ik^{\prime}y}dk^{\prime}.
\]
Using $\int_{-\infty}^{\infty}dt\exp[-i(k^{\prime}\pm k)v_{g}t]=\frac{2\pi
}{v_{g}}\delta(k^{\prime}\pm k)$, $E_{B}$ is estimated as
\begin{align}
E_{B}  &  =\frac{3e^{3}vR\nu_{S}}{4\pi^{3}\epsilon\Delta V}\int_{a_{-}}%
^{a_{+}}d\bar{x}_{A}\int_{b_{-}}^{b_{+}}d\bar{y}_{B}\int_{b_{-}}^{b_{+}}%
dx_{B}\int_{b_{-}}^{b_{+}}dy_{B}\nonumber\\
&  \times\frac{1}{\sqrt{\left(  x_{B}-y_{B}\right)  ^{2}+d^{2}}}\\
&  \times\frac{w_{A}(\bar{x}_{A})\lambda_{B}(\bar{y}_{B}-L)}{\left(
x_{B}+y_{B}-\bar{x}_{A}-\bar{y}_{B}+L+v_{g}T\right)  ^{5}},
\end{align}
where $v_{g}T=O(10^{-2}L)$. The parameter $L+v_{g}T(=O(L))$ corresponds to the
distance between A and B. Thus, the energy output $E_{B}$ is estimated as%
\begin{equation}
E_{B}\sim\frac{e^{2}\lambda_{B}}{4\pi\epsilon l}\frac{ev_{g}R}{l\Delta
V}\left(  \frac{l}{L}\right)  ^{5}. \label{Eb}%
\end{equation}
It should be emphasized here that a positive function $\lambda_{B}(\bar{y}%
_{B}+L)$ guarantees positive $E_{B}$. Obviously from Eq. (\ref{Eb}), an
increase in $L$ rapidly degrades the magnitude of $E_{B}$ (e.g., $E_{B}\sim
$1$~\mathrm{\mu}$eV for $L\sim4l$). Nevertheless, for $L\sim2l$, $E_{B}%
\ $attains a value on the order of $100~\mathrm{\mu}$eV. This is much larger
than the thermal energy $\sim1~\mathrm{\mu}$eV at a temperature of $\sim
10~$mK, at which experiments on the QH effect are often performed (using a
dilution refrigerator). Note here that to estimate actual value of $E_{B}$, we
need to know $E_{1}$ since the energy, which can be measured by the setup in
Fig.~\ref{fig:model}, is $E_{2}$ ($=E_{B}+E_{1}$). $E_{1}$ can be estimated by
letting $d$ be sufficiently large\cite{device}.

To observe $E_{B}$ experimentally, we turn on the switch and measure the
current passing through the edge channel $P$ once (single-shot measurement).
The relation
\[
\varepsilon=\frac{\pi\hbar}{\nu_{P}e^{2}v_{g}}j^{2},
\]
between the energy density $\varepsilon$ and current $j$ gives an energy
density of $10$-$\mathrm{\mu eV/\mu m}$, which corresponds to a current of
$10$-$\mathrm{nA}$. This current can be detected experimentally using a
picoammeter. To verify that energy is extracted at B, a sufficient number of
single-shot current measurements should be conducted (by switching the circuit
on and off) to generate meaningful statistics for the POVM measurement. In
this process, the electrical noise, which can be introduced in the classical
channel, is averaged out and thus does not affect $\langle E_{B}\rangle$. Note
here that $(l/L)^{5}$ dependence of the estimated $E_{B}$ is based on the
first-order perturbation theory and the dependence might be slower than in
higher-order approximations or in a framework of more suitable local
operations. Careful discussion is need for optimizing the experimental setup
to obtain maximum $E_{B}$.

\section{\label{sec4}Discussion and Conclusion}

We now examine energy conservation and dynamics in the system. As we have
shown, the extraction of $E_{B}$ from the local vacuum state requires
measurement (energy injection) at A. What is the source of $E_{A}$? We
consider a POVM measurement, so that switching on the RC circuit causes energy
$E_{A}$ to be injected into $S$. Therefore, if the switch is electrically
operated, a battery may provide $E_{A}$ to drive the switching device
\cite{switchU}. After extracting $E_{B}$, the total energy $E_{A}-E_{B}$ of
the system will be non-negative, as expected, because $E_{A}>E_{B}$. According
to the local energy conservation laws, the transfer of energy $E_{B}$ from $S$
to $P$ results in a negative average quantum energy density around B. This
negative energy density is obtained by squeezing the amplitude of the
zero-point fluctuation to less than that of the vacuum state during the
interaction \cite{negativeE}. Then, $-E_{B}$ and $E_{A}$ will flow
unidirectionally along the edge toward the downstream electrical ground with
identical velocities of $v_{g}$, and $S$ around region B will remain in a
local vacuum state with zero energy density.

Although no studies have been conducted on QET in QH systems, several
successful experimental studies have been conducted in quantum optics by
introducing LOCC including QT \cite{QT,exp-QT}. Light is a massless
electromagnetic field; however, at present, it is difficult to directly
measure the zero-point fluctuations of light owing to the lack of an
appropriate interaction such as the Coulomb interaction in QH systems. Thus,
our QH system is considered to be very suitable for demonstrating the QET protocol.

QET can be interpreted in terms of information thermodynamics as a quantum
version of Maxwell's demon \cite{demon}; in particular, two demons
cooperatively extract energy from quantum fluctuations at zero temperature.
Moreover, this type of quantum feedback is relevant to black hole entropy,
whose origin has often been discussed in string theory \cite{strominger},
because energy extraction from a black hole reduces the horizon area (i.e.,
the entropy of the black hole \cite{bh}).

In conclusion, we have theoretically shown the implementation of QET and
estimated the order of the energy gain\ $E_{B}$ in a QH system using
reasonable experimental parameters.

\begin{acknowledgments}
The authors thank K. Akiba and T. Yuge for the fruitful discussions. G. Y., W.
I., and M. H. are supported by Grants-in-Aid for Scientific Research (Nos.
21241024, 22740191, and 21244007, respectively) from the Ministry of
Education, Culture, Sports, Science and Technology (MEXT), Japan. W. I. and M.
H. are partly supported by the Global COE Program of MEXT, Japan. G. Y. is
partly supported by the Sumitomo Foundation.
\end{acknowledgments}


\begin{thebibliography}{99}
\bibitem {casimir}In fact, this energy is responsible for the Casimir force
between two parallel conducting plates; however, it is impossible to locally
extract this force from the vacuum state without changing the boundary
conditions of the field such as in the Casimir effect.
\bibitem {chirarity}It should be stressed that since the edge current of $S$
flows only to the left and that the upstream of subsystem~B is connected to an
ideal electric ground, the zero-point fluctuations at subsystem B are not
affected by events that occur in the downstream regions. Thus, subsystem B
always remains in the local vacuum state before the local operation at B.
\bibitem {impedance}The characteristic impedance of wires is matched to $R$.
\bibitem {device}For example, we place a metal gate electrode between the two
edge channels. By applying a negative voltage to the gate, the two-dimensional
electron density can be locally depleted. In this fashion, the effective
distance $d$ between the two edge channel can be continuously controlled. Such
a technique is commonly used in experiments related to semiconductor nanostructures.
\bibitem {switchU}Unlike the local measurement, which can be connected to $S$
by the switch only during the measurement, the local operation is always
attached to $S$. In the presented analysis, we can omit the modification
effects induced by the existence of $U$ adjacent to $S$ because such a
backaction effect appears only in higher correction terms of the pertubation.
\bibitem {negativeE}Such emergence of negative energy density is widely
discussed in quantum field theory. For example, one of the most simple cases
involves the linear superposition of the vacuum state and the multi-particle
states of a quantum field \cite{Ford}.
\end{thebibliography}
\end{document}